\documentclass[12pt]{article}
\usepackage{graphicx}

\begin{document}

\setcounter{page}{0}
\begin{titlepage}
\title{Fermionic Ising glasses in magnetic transverse field with BCS pairing interaction}
\author{S. G. Magalh\~aes and F. M. Zimmer\\[3mm]         
Departamento de Fisica\\      
Universidade Federal de Santa Maria\\         
97105-900 Santa Maria, RS, Brazil\\}
\date{}
\maketitle
\thispagestyle{empty}
\begin{abstract}
\normalsize
\noindent 

We study a fermionic infinited-ranged Ising spin glass with a real space BCS interaction  
in the presence of an applied transverse field. The problem is formulated in the integral 
functional formalism where the $SU(2)$ spins are given in terms of bilinear combinations of Grassmann fields. 
The problem is solved within static approximation and the replica symmetry ansatz combined with previous approaches 
used  to study the critical behavior of the quantum Ising spin glass in a transverse field and the  spin glass 
Heisenberg model with BCS pairing. 
Our results show that the transverse field has strong effect in the phase boundary of the spin glass phase and the 
PAIR phase in which there is a long range order corresponding to formations of pairs.  The location of the tricritical 
point in the PAIR phase transition line is also affected.      

\end{abstract}
\vspace*{0.5cm}
\end{titlepage}

\setcounter{page}{1}
\section{Introduction}
Theoretical studies in recent years have been investigating the interplay between superconductivity (SC) and 
spin glass (SG) \cite{Castro-Neto,magal99,magal00,Oppermann1,Galitski} which has been found in several stron\-gly 
correlated electron systems, such as heavy fermions \cite{1} and cuprate superconductors \cite{2}. 
However, the experimental evidences for these correlated systems have 
showed a quite complex situation. For instance, the heavy fermion superconductor 
$U_{1-x}$$M_{x}$$Pd_{2}Al_{3}$ ($M=La$, $Y$, $Th$) \cite{1} shows a sequence of magnetic and Non-Fermi Liquid (NFL)  gro\-und states. 
In particular, when the content of $La$ is increased, a antiferromagnetic phase is replaced by a SG which is suppressed to 
$T=0K$ at $x=0.75$. After the Quantum Critical Point (QCP), there is NFL region and at $x=1$ appears superconductivity. 

Some approaches have proposed  that disorder itself can be the source of deviation of 
the Fermi Liquid behavior. For heavy fermions, 
for instance, the so called Kondo disordered model (KDM) 
\cite{Miranda} describes a distribution of Kondo temperature $T_{K}$
given origin to the NFL behaviour. 
Castro-Neto et al. \cite{Castro} relates the 
NFL effects to the presence of an inhomogeneous Griffith's phase. 
In an earlier work, it has been shown a NFL behavior near to the QCP.
in a transition between a metallic-paramagnet and a metallic-spin glass \cite{Georges}. But,  there is also the suggestion that 
the presence of QCP itself can be a source of a new class of excitations which spread its effects  even at finite temperatures 
leading to a breakdown of the Fermi liquid theory \cite{ColeMu}.  

However, little consideration has been given to describe how the phase boundaries between superconductivity and 
spin glass are modified when at low temperature, besides the presence 
of thermal fluctuations, the quantum fluctuations start to become important. Some of the models previously mentioned
\cite{magal99,Oppermann1} have addressed the phase transition problem between SG and superconductivity. Nevertheless, in those 
references there is no mechanism able to tune quantum fluctuations.     

In particular, the model in Ref. \cite{magal99}  
has been derived from a model introduced by Nass et al \cite{Nass} to deal with conventional superconductors doped with magnetic 
impurities. This model is an s-d exchange interaction  between the magnetic impurities together with a conventional BCS interaction 
between the conducting electrons.  
When the conduction electrons are integrated by a perturbation expansion to second order in $J_{sd}$ (the exchange interaction), 
the resulting effective model consists of the RKKY interaction with a pairing interaction between localized fermions. 
If the coupling $J_{ij}$ between the localized magnetic moments is assumed to be random gaussian distributed, one has 
a model to study the phase boundaries between SG and 
a phase where there is spin pairing \cite{magal99}. 

The functional integral machinery, where the SU(2) spins have been represented by bilinear combination  of the Grassmann variables, 
has been the suitable method used \cite{magal99} 
to study the mentioned effective model in its Ising version.  Combined with replica trick and the static approximation, 
it has allowed to find the Grand-Canonical potential in terms of the spin glass and the PAIR order parameters. 
The last phase corresponds to a long range order where there is pair formation. In the half-filling situation, 
the results have showed a phase diagram temperature $T/J$ {\it versus} $g/J$ ($g$ is the strength of the pairing interaction and
$J$ is the variance of $J_{ij}$) 
where one can find a spin glass phase for low temperature and small $g$. If $g$ is increased, one gets a phase 
transition at $g=g_{1}(T)$ where there is a PAIR phase. The nature of the transition line is complex presenting a 
tricritical point ($T_{tc}, g_{tc}$). 

Further investigation studied the pairing-spin glass competition replacing the fermionic Ising in the cited effective model by the  
Heisenberg model with an applied magnetic field $H_{z}$ within the same framework of References \cite{magal99} and \cite{Gold}. 
The results have showed that the transition to the PAIR phase depends on the replica diagonal spin glass order 
parameter (which is associated with the susceptibility) even at higher temperatures than the freezing temperature $T_{f}$. 
The region in temperature where the calculated line transition 
between the normal-paramagnetic (NORMAL) and the PAIR phases is first order  
becomes larger when $H_{z}$ is increased. Nevertheless, the line transition between the SG and the 
PAIR phase have not been accessible in this work.  
Other interesting point about the infinite ranged 
quantum Heisenberg model in the presence of a field is that $T_{f}$ is depressed 
when the field increases, but never reaches a QCP \cite{magal00}.       

Corrections in the weak hopping 
limit to the Ising SG fermionic model in the presence 
of a local pairing interaction have been studied elsewhere \cite{Oppermann1}. The results show that those corrections essentially 
preserves the shape of the phase 
diagram obtained with no hopping. Therefore, one can consider the 
the range of validity of this theory as covering the transition between poor conductors 
and superconductors.   

Recently, the Ising SG alone has been investigated in the transverse field $\Gamma$.  
The functional integral approximation have been used \cite{Alba2} to deal with the non-commutativity of the spins operators 
which have been represented by bilinear combinations 
of Grassmann fields. There are two versions of SG fermionic problem.   The first one, the 
operator $S_{i}^{z}$ has four eigenvalues (two of them are non-magnetic). In the second model, 
the two vanishing eigenvalues are suppressed by a restraint. In both models, the freezing temperature decreases  
with increasing $\Gamma$ until  to reach a QCP at a critical value $\Gamma_{c}$ 
This sort of approach is a natural tool to study phase transitions in condensed matter 
problems where  fermions experiment couplings such as superconductivity and Kondo effect \cite{Coqblinalba}. 

Therefore, our aim in this work have been to investigate how the  phase boundary between a PAIR phase 
(where there is pair formation) and the four-state fermionic spin glass (SG) is modified 
if there is spin flipping induced by the presence of the transverse field $\Gamma$. Therefore, 
the transverse field plays the role of the spin flipping part of Heisenberg model \cite{Alba3} allowing to access the QCP.
In order to solve the functional integral 
over the Grassmann fields contained in the partition function, 
 the formalism of Nambu matrices and spinors has been used.
 We also have used the replica symmetric ``ansatz", and therefore we have calculated the Almeida-Thouless 
line \cite{refat} to obtain the validity limit of this procedure.
Finally, we find the  Grand-Canonical potential and the saddle point equations for the order parameters in the half-filling limit. 

One important approximation in the present work is the neglecting of time fluctuations (the static approximation) \cite{Bray}. 
For infinite ranged quantum Ising spin glass, a simulational approach \cite{Grempel} has shown that for temperatures close 
to the freezing temperature $T_{f}$ (when $\Gamma=0$) the static approximation can be considered reliable. On the other hand, 
it is quite clear that the static approximation 
is unable to capture the fundamental low temperature dynamical behavior of the correlation functions \cite{Huse}. 
Nevertheless, it has been shown that for $M$-component quantum rotor model, in the limit $M=\infty$, 
the critical line is given by zero-frequency mode \cite{Sachato}. This critical behaviour coincides  with the 
Ising SG in transverse field \cite{Huse}.  
Therefore, that is the ultimate justification for the use of the static approximation to find the phase boundary between SG and the 
PAIR phase for increasing transverse field which is the main purpose of the present work.  

This paper is organized as follow. In section II, 
we introduce the model and perform calculations using the replica trick and the static approximation 
in order to find the Grand canonical potential and the saddle point order parameters equations. The behavior of 
the tricritical point in the transition line to the PAIR phase is obtained as a function of both $\Gamma$ and $g$.  
In section III, phase diagrams are build up 
with solutions from the set of the order 
parameter equations in both situations, $T/J$ {\it versus} 
$g/J$ (for two values of $\Gamma$) and $T/J$ {\it versus} 
$\Gamma/J$ (for several values of $g$). It is also suggested a relationship between $\Gamma$ and $g$ 
which allow to see more clearly the role of quantum fluctuations in the interplay between SG and the PAIR phase. 
In the last section, we present our conclusions and final remarks.          

\section{General Formulation}

The model considered in this work was obtained by tracing out the conducting
electrons degrees of freedom of a superconductor alloy \cite{magal99}, resulting
in an effective BCS pairing interaction among fermions and a random Gaussian 
interaction coupling the localized spins. In the resulting effective model we apply the transverse field term.  
Therefore, the Hamiltonian is 
\begin{eqnarray}
\hat{H}-\mu \hat{N}=-\sum_{i,j} \left[J_{ij}\hat S_{i}^{z} \hat S_{j}^{z}+
\frac{g}{N}c_{i\uparrow}^{\dagger}c_{i\downarrow}^{\dagger}c_{j\downarrow}c_{j\uparrow}\right]
\nonumber \\-
\sum_{j}
2\Gamma \hat S_{j}^{x}+\sum_{j} \mu\sum_{s=\uparrow,\downarrow} \hat{n}_{js}
\label{ham}
\end{eqnarray}
where the sum is over the $N$ sites of a lattice. The coupling $J_{ij}$ is an independent random variable 
with Gaussian probability distribution given by
\begin{eqnarray}
P(J_{ij})=\sqrt{\frac{N}{32\pi J^2}}\exp\left(-\frac{J_{ij}^2}{32 J^2/N}\right)
\label{gaussian}~.
\end{eqnarray}

The spin operators in Eq. (\ref{ham}) are defined (see references \cite{Alba2,magal99}) as:
\begin{eqnarray}
\hat S_{j}^{z}=\frac{1}{2}[\hat{n}_{j\uparrow}-\hat{n}_{j\downarrow}]
~;~~~\hat S_{j}^{x}=\frac{1}{2}[c_{j\uparrow}^{\dagger}c_{j\downarrow}
+c_{j\downarrow}^{\dagger}c_{j\uparrow}]
\label{opex}
\end{eqnarray}										
where the $c_{j\sigma}^{\dagger}~(c_{j\sigma})$ are fermions creation (destruction) operators, w
ith $\sigma=\uparrow$ or $\downarrow$ indicating the spin projections, $\hat{n}_{j\sigma}=c_{j\sigma}^{\dagger}c_{j\sigma}$ is the 
number operator and $\mu$ is the chemical potential.   

The Grand Canonical partition function is formulated in the functional integral 
formalism for fermions using the anticommuting Grassmann variable
$\phi_{js}^{*}(\tau)$ and $\phi_{js}(\tau)$( $\tau$ is the complex time). Therefore one has
\begin{eqnarray}
{\cal Z}=\int D(\phi^{*}\phi) ~\exp \left[ A_0+A_{SG}+A_\Gamma+A_{BCS}\right]
\label{can}
\end{eqnarray} 
where the actions $A_0$, $A_{SG}$, $A_{\Gamma}$, and $A_{BCS}$ are the free part, the spin glass part, 
the transverse field part and the pairing part, respectively.  The three first ones assume (after time Fourier transformation) 
the following forms: 
\begin{eqnarray}
A_{0}=\sum_{j}\sum_{\omega}\underline{\phi}_{j}^{\dag}(\omega)(i\omega+\beta\mu)
\underline{\phi}_{j}(\omega)
\label{eqao},
\\
A_{_{SG}}=\sum_{ij}\beta J_{ij}S_{i}^{z}(\Omega)S_{j}^{z}(-\Omega)
\label{eqasg},
\\
A_{\Gamma}= \sum_{j}\sum_{\omega}\beta \Gamma
\underline{\phi}_{j}^{\dag}(\omega)
\underline{\sigma}_{1}\underline{\phi}_{j}(\omega)
\label{eqagamma}
\end{eqnarray}
where
$S_{j}^{z}(\Omega)=\frac{1}{2}\sum_{\omega}\underline{\phi}_{j}^{\dagger}(\omega+\Omega)
\underline{\sigma}_3 \underline{\phi}_{j\sigma}(\omega)$, 
with Matsubara's frequencies $\omega=(2m+1)\pi$ and $\Omega=2m\pi$ ($m=0,\pm
1,\cdots$). In the Eqs. (\ref{eqao})-(\ref{eqagamma}) we have used the
Spinors
\begin{equation}
\underline{\phi}_{j}(\omega)=\left[\begin{tabular}{c}$\phi_{j\uparrow}(\omega)$ \\ 
$ \phi_{j\downarrow}(\omega)$ \end{tabular}
\right]; \ \ \ \underline{\phi}_{j}^{\dag}(\omega)=\left[\begin{tabular}{c}
$\phi_{j\uparrow}^{*}(\omega)$ \ $\phi_{j\downarrow}^{*}(\omega)$\end{tabular}
\right]
\label{spinor}
\end{equation}
and the Pauli matrices
\begin{equation}
\underline{\sigma}_{1}=\left(
\begin{tabular}{cc}
$0$ & $1$\\  $1$ & $0$
\end{tabular}
\right);~ 
\underline{\sigma}_{2}=\left(
\begin{tabular}{cc}
$0$ & $-i$\\  $i$ & $0$
\end{tabular}
\right);~ 
\underline{\sigma}_{3}=\left(
\begin{tabular}{cc}
$1$ & $0$\\  $0$ & $-1$
\end{tabular}
\right).
\label{pauli}
\end{equation}
The pairing action is given by
\begin{equation}
A_{BCS}=\sum_{ij}\sum_{\Omega}\rho_{i}^{*}(\Omega)\rho_{j}(\Omega)
\label{eqabcs}
\end{equation}
with
$\rho_{j}(\Omega)=\sum_{\omega}\phi_{j\downarrow}(-\omega)\phi_{j\uparrow}(\Omega+\omega)$.

In this paper we discuss the phase transition problem within the static approximation, 
therefore only the term with $\Omega=0$ is kept in the sum over the Matsubara`s frequencies 
in Eqs.(\ref{eqasg}) and (\ref{eqabcs}). 
For this reason, we can define  $S_j\equiv S_j^z(0)$ and write the following expression for 
the pairing action
\begin{equation}
A_{_{BCS}}^{st}=\frac{\beta g}{4N}\sum_{p=1,2}\left[\sum_{j,\omega}
\underline{\phi}_{j}^{'\dag}(\omega)\underline{\sigma}_{p}\underline{\phi}^{'}_{j}
(\omega)\right]^{2},
\label{pares}
\end{equation}
where the Nambu matrices have been introduced in the previous equation, 
\begin{equation}
\underline{\phi}_{j}^{'}(\omega)=\left[\begin{tabular}{c}$\phi_{j\uparrow}(\omega)$ \\
 $\phi_{j\downarrow}^{*}(-\omega)$\end{tabular}\right] ;~
\underline{\phi}_{j}^{'\dag}(\omega)=\left[\begin{tabular}{c}$\phi_{j\uparrow}^{*}(\omega)$ 
\ $\phi_{j\downarrow}(-\omega)$\end{tabular}\right]
\label{nambu}.
\end{equation}

The configurational averaged Grand Canonical potential per site can be found by using the replica formalism
\begin{eqnarray}
\frac{\Omega}{N}=-\frac{1}{N\beta}\lim_{n\rightarrow 0}\frac{{\cal{Z}}(n)-1}{n}
\label{truque}.
\end{eqnarray}
The configurational averaged replicated partition function 
${\cal Z}(n)=\langle{\cal{Z}}^{n}\rangle_{J_{ij}}$ becomes, after averaging 
over $J_{ij}$, 
\begin{eqnarray}
{\cal Z}(n)=\int D(\phi_{\alpha}^*,\phi_{\alpha}) \exp\left\{
\sum_{\alpha}
[A_0^\alpha+A_\Gamma^\alpha]+
\right.
\nonumber \\ \left.
\frac{\beta g}{4 N}\sum_{\alpha}\sum_{p=1,2}
\left(\sum_{j,\omega}\underline{\phi}_{j}^{' \alpha\dagger}(\omega)
\underline{\sigma}_p \underline{\phi}_{j}^{' \alpha}(\omega)
\right)^2 
\right.
\nonumber \\ \left.
+\frac{8\beta^2J^2}{N}\sum_{\alpha,\beta}
\left(\sum_{j}S_j^\alpha S_j^\beta
\right)^2 \right\}.
\label{zn1}
\end{eqnarray}

In the previous equation we introduce the replica index $\alpha=1,2,\cdots,n$. 
The linearization of Eq. (\ref{zn1}) is obtained by using the Hubbard-Stratonovich transformation 
\begin{eqnarray}
{\cal Z}(n)={\cal{N}}\int_{-\infty}^{\infty}\prod_{\alpha}d\eta_{R\alpha}d\eta_{I\alpha}
\int_{-\infty}^{\infty}\prod_{\alpha\beta}dq_{\alpha\beta}
\exp
\left\{-N \right.\nonumber\\ 
\times(\beta g\sum_{\alpha}|\eta_{\alpha}|^{2}
+\frac{\beta^{2}J^{2}}{2}
\sum_{\alpha\beta}q_{\alpha\beta}^{2}-
\ln\Lambda_{\alpha}(q_{\alpha\beta},\eta_\alpha) )
\}
,~
\label{zn2}
\end{eqnarray}
where was introduced replica dependent auxiliary fields $|\eta_\alpha|$ and
$q_{\alpha\beta}$. In the Eq. (\ref{zn2}) $\eta_{\alpha}=\eta_{R\alpha} - i
\eta_{I\alpha}$, ${\cal N}=(\beta g N/\pi)^n(\beta^2 J^2 N/ 2\pi)^{n}$, where
\begin{eqnarray}
\Lambda_{\alpha}(q_{\alpha\beta},\eta_\alpha)=\int \prod_{\alpha=1}^{n}D[\phi^{\alpha *}\phi^{\alpha}]
\exp\left[\sum_{\alpha}\left(A_{0}^{\alpha} + A_{\Gamma}^{\alpha}\right) 
\right.\nonumber\\
+ 4\beta^{2}J^{2}\sum_{\alpha\beta}q_{\alpha\beta}S_{}^{\alpha}S_{}^{\beta}
\left.+\sum_{\omega}
\underline{\phi}^{'\dag\alpha}(\omega)
\underline{\eta}_{\alpha}\ \underline{\phi}_{}^{'\alpha}(\omega)
\right],~
\label{particao3}
\end{eqnarray}
with the matrix $\underline{\eta}_\alpha$ defined as:
\begin{equation}
\underline{\eta}_{\alpha}=
\left(\begin{tabular}{cc}
$0$ & $\beta g\eta_{\alpha}$\\  $\beta g \eta_{\alpha}^{*}$ & $0$
\end{tabular}\right)\ .
\label{e33}  
\end{equation}
The order parameter $|\eta_\alpha|$
introduced in the Eq. (\ref{zn2}) corresponds to a long range order where there 
is double occupation of the sites \cite{magal99}, and $q_{\alpha\beta}$ is the spin glass order parameter. 

In the present work we restrict the discussion to the replica symmetric ansatz, 
that considers
\begin{eqnarray}
q_{\alpha \beta}= q; ~~q_{\alpha \alpha}= q + \bar{\chi}
\label{e42} 
\end{eqnarray} 
where $q$ is the spin glass order parameter and $\bar{\chi}=\frac{\chi}{\beta}$ 
($\chi$ is the static susceptibility \cite{albagusmao}).
One can sum over the replica indices, which produces new quadratic terms 
that are linearized by introducing new auxiliary fields. Therefore,
the functional integral becomes
\begin{eqnarray}
\Lambda_{\alpha}(q_{\alpha\beta},\eta_\alpha)=\int_{-\infty}^{\infty}Dz
\left[\int_{-\infty}^{\infty}D\xi {\cal I}(\xi,z,h)\right]^n
\label{lambda}
\end{eqnarray} 
with $~Dz=dz\frac{\mbox{e}^{-z^2/2}}{\sqrt{2\pi}}$, 
$~D\xi=d\xi\frac{\mbox{e}^{-\xi^2/2}}{\sqrt{2\pi}}~$, and
\begin{eqnarray}
{\cal I}(\xi,z,h)=\int D[\phi^{*}\phi]\exp\left[
\sum_{\omega}\underline{\phi}^{\dag}(\omega)
 \underline{G}_{1}^{-1}(\omega) \underline{\phi}(\omega)
\right.\nonumber\\
\left.+
\sum_{\omega}
\underline{\phi}^{'\dag}(\omega)
\ \underline{\eta}\ \underline{\phi}^{'}(\omega)\right]
\label{e49}
\end{eqnarray} 
where the matrix  $\underline{G}_{1}^{-1}(\omega)$ is defined by
$\underline{G}_{1}^{-1}(\omega_n)=i\omega_n +\beta\mu +\beta\Gamma \underline{\sigma}_1
 +h\underline{\sigma}_3~$ and the field $h=\beta J\sqrt{2\bar{\chi}}\xi+
 \beta J\sqrt{2q}z$.

In order to solve the integral in Eq. (\ref{e49}) which combines the
elements of spinors and Nambu matrices, we can use a similar 
procedure already done in Ref. \cite{magal00} 
which mixes the elements of  
the spinors and the Nambu matrices  to write 
Eq. (\ref{e49}) as:
\begin{eqnarray}
{\cal I}(\xi,z,h)=\int D[\phi^{*}\phi]\exp\left[
\sum_{\omega}\underline{\Phi}^{\dag}(\omega)
 \underline{G}^{-1}(\omega) \underline{\Phi}(\omega)\right]
 \label{e50}
\end{eqnarray}  
where 
\begin{eqnarray}
\underline{\Phi}^{\dag}(\omega)=\left [\begin{tabular}{cccc}
$\phi_{\uparrow}^{*}(\omega)$ \ $\phi_{\downarrow}^{*}(\omega)$  \ 
$\phi_{\downarrow}(-\omega) $\ $ \phi_{\uparrow}(-\omega)$\end{tabular}\right ]
\label{e51}
\end{eqnarray}
and
\begin{equation}
\underline{G}^{-1}(\omega)=\left(
\begin{tabular}{cccc}
$i\omega+\zeta^+$ & $\beta\Gamma$ & $\beta g\eta$ & $0$\\
\\$\beta\Gamma$ & $i\omega+\zeta^-$ & $0$ & $-\beta g\eta$\\
\\$\beta g\eta^{*}$ & $0$ & $i\omega-\zeta^+$ & $-\beta\Gamma$\\
\\$0$ & $-\beta g\eta^{*}$ & $-\beta\Gamma$ & $i\omega-\zeta^-$
\end{tabular}
\right)
\label{e53}
\end{equation} 
with ${\zeta^{\pm}}=\beta\mu \pm h$.

In the Eq. (\ref{e50}), the differential $D[\phi^{*}\phi]$ stands 
for $\prod_{\omega}$ $\prod_{\sigma=\uparrow\downarrow}$ $d\phi^{*}_\sigma(\omega)$ $d\phi^{*}_\sigma(-\omega)$ $d\phi_\sigma(\omega)
$ $d\phi_\sigma(-\omega)$. 
The functional integral over the Grassmann fields and the sum over 
the Matsubara frequencies can be readily performed with the result \cite{magal99,magal00}: 
\begin{eqnarray}
{\cal I}(\xi,z,h)=\cosh\sqrt{(\beta\mu)^{2}+(\beta g |\eta|)^2} +
\cosh\sqrt{\Theta}
\label{e56}
\end{eqnarray}
where $\Theta=h^{2}+(\beta\Gamma)^{2}$.

The results obtained in Eq.(\ref{e56}) can be used in the  Eq. (\ref{lambda}) 
that allow us to rewrite the Eq. (\ref{zn2}). Therefore, the saddle point 
method  (see Eq. (\ref{truque})) give us the Grand Canonical potential as   
\begin{eqnarray}
\frac{\beta \Omega}{N}=\beta g \eta^2
+ \frac{\beta^2 J^2}{2}\bar{\chi}(\bar{\chi}+2q)
-\int_{-\infty}^{\infty}
Dz\ln I_a(z)
\label{potencial}
\end{eqnarray}
where 
\begin{eqnarray}
I_a(z)=\cosh{\beta g\eta}+\int_{-\infty}^{\infty}D\xi
\cosh{\sqrt{\Theta}}
\label{complem}
\end{eqnarray}
with the chemical potential fixed to ensure that we are in the half-filling situation. 
From now, the parameter $\eta$ is used instead of $|\eta|$. The saddle point equations for order parameters 
that follow from equation (\ref{potencial}) are:
\begin{eqnarray}
\eta=\frac{1}{2}\int_{-\infty}^{\infty}Dz \frac{\sinh(\beta g \eta)}
{I_a(z)}
\label{eta}~,
\\
q=\int_{-\infty}^{\infty}Dz\left[
\frac{\int_{-\infty}^{\infty}D\xi\ h\ \sinh\sqrt{\Theta}}
{I_a(z)}\right]^{2}
\label{q}~,
\\
\bar{\chi}=\int_{-\infty}^{\infty}Dz
\frac{I_b(z)}{I_a(z)}-q~,
\label{chi}
\end{eqnarray}
where
\begin{eqnarray}
I_b(z)=\int_{-\infty}^{\infty}D\xi
\left( \frac{\beta^{2}\Gamma^{2}\sinh\sqrt{\Theta}}
{\Theta^{\frac{3}{2}}}
+\frac{h^{2}\cosh\sqrt{\Theta}}{\Theta}
\right).
\label{Int}
\end{eqnarray}

The solution with replica symmetric is unstable at low temperature. 
Therefore, it is necessary to find the region in temperature where the Almeida-Thouless 
eigenvalue \cite{refat} $\lambda_{AT}$ becomes negative, which is given by:  
\begin{eqnarray}
\lambda_{AT}=1-2(\beta J)^2\int_{-\infty}^{\infty}Dz
\left(\frac{I_a(z)I_b(z)-I_c^2}{I_a(z)}\right)^2
\label{almeida}.
\end{eqnarray} 
In the Eq. (\ref{almeida}) $I_c(z)$ is
\begin{eqnarray}
I_c(z)=\int_{-\infty}^{\infty}D\xi
\frac{h\sinh\sqrt{\Theta}}
{\sqrt{\Theta}}
.
\label{AT}
\end{eqnarray}

The Landau expansion of the Grand Canonical potential in powers of the two 
parameters $\eta$ and $q$ allow us  to locate the second order line transition in the problem. 
Therefore, from Eqs. (\ref{potencial})-(\ref{complem}): 
\begin{eqnarray}
\beta \Omega=\sum_{j=0}^{3}f_{j}(\eta,\bar{\chi},\beta)q^j 
\label{expansao1}
\end{eqnarray}   
where $\bar{\chi}(q,\eta,\beta)$ is the solution of the saddle point equation,
that was rewrite in powers of $q$ as: 
\begin{eqnarray}
\bar{\chi}=\bar{\chi}_0+\bar{\chi}_1q+\bar{\chi}_2q^2
\label{expansao2}.~
\end{eqnarray}  

Introducing the Eq. (\ref{expansao2}) into Eq. (\ref{expansao1}), and expanding
the coefficients $f_j$ in powers of $q$ and $\eta$, we obtain the following
result:
\begin{eqnarray}
\frac{\beta \Omega}{N}=\frac{\beta^2
J^2}{2}\bar{\chi}^2_0-\ln(K_0)+A_2q^2+A_3q^3
\nonumber\\
+B_2\eta^2 + B_4\eta^4 + B_6\eta^6
\label{A1}
\end{eqnarray} 
with 
\begin{eqnarray}
A_2=-\frac{\beta^2J^2}{2!}+ \beta^4J^4 \bar{\chi}^2,
~~A_3=-\frac{8\beta^6J^6}{3}\bar{\chi}^3_0,
\label{A3}
\\
B_2=\beta g - \frac{\beta^2g^2}{2!K_0},
\label{B2}
\\
B_4=-\frac{\beta^4 g^4}{4!K_0^2}
\int_{-\infty}^{\infty}D\xi(\cosh\sqrt{\Delta}-2),
\label{B4}
\\
B_6=-\frac{\beta^6 g^6}{6!K_0^3}\left[ 16-
13\int_{-\infty}^{\infty}D\xi\cosh\sqrt{\Delta}
\right.\nonumber\\
\left.+\left(\int_{-\infty}^{\infty}D\xi\cosh\sqrt{\Delta}\right)^2\right]
\label{B6}
\end{eqnarray}
where $K_0=1+\int_{-\infty}^{\infty}D\xi\cosh\sqrt{\Delta}~$,
$~\bar{\chi}_0=\frac{1}{K_0}\int_{-\infty}^{\infty}D\xi$ $\frac{\sinh{\sqrt{\Delta}}}{\sqrt{\Delta}}~$,
 and $~\Delta=2\beta^2J^2\bar{\chi}_0\xi^2+\beta^2\Gamma^2$.
\begin{figure}[t]
\includegraphics[angle=270,width=12cm]{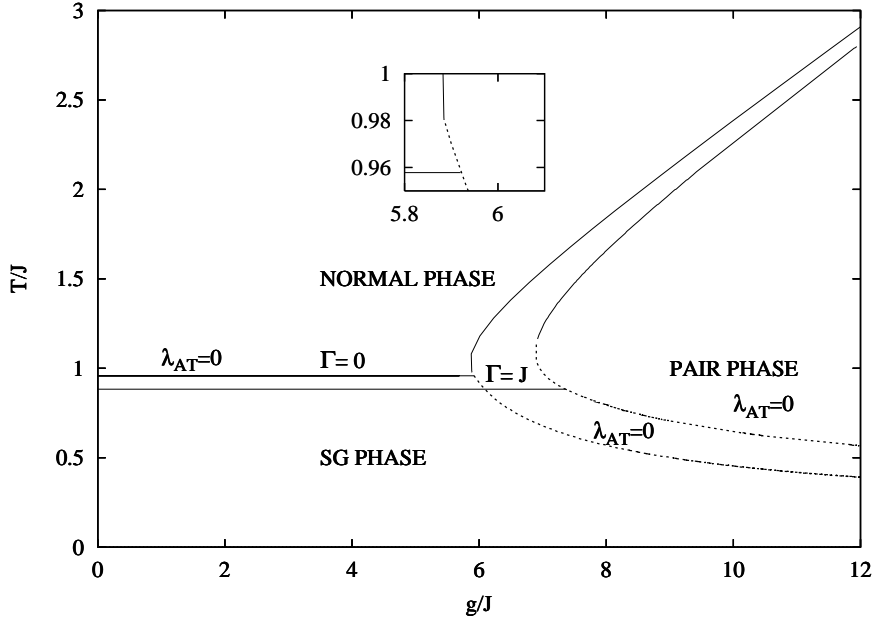}
\caption{Phase diagrams as a function of $T/J$ and pairing coupling $g/J$ for two
values of $\Gamma$. Solid lines indicate second-order transition while dotted
line indicate a first-order transition. The tricritical point for $\Gamma=0$ is 
shown in detail in the diagram.}  
\label{fig1}
\end{figure}

\begin{figure}[t]
\includegraphics[angle=270,width=12cm]{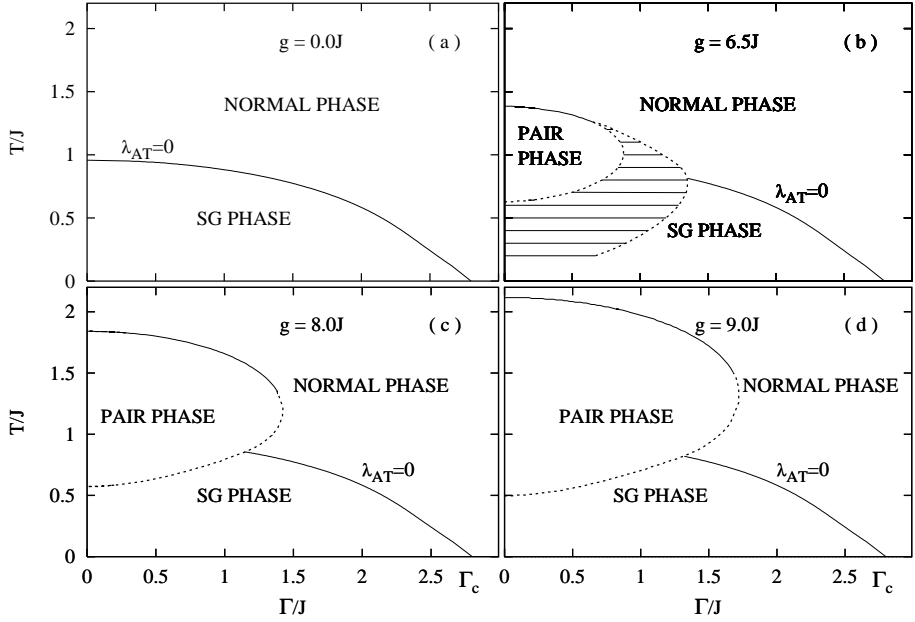}
\caption{Phase diagrams as a function of $T/J$ and $\Gamma/J$ for several fixed 
values of 
$g/J$: (a) $g=0$, (b) $g=6.5J$, (c) $g=8J$, and (d) $g=9J$. It is used 
the same convention as Fig. (\ref{fig1}) for the transition lines. 
The hatched region in panel (b) delineates a multiple solution region.}  
\label{fig2}
\end{figure}

The tricritical point is given when both coefficients $B_2$ and $B_4$ change the
sign. In this condition we have from Eqs. (\ref{B2}) and (\ref{B4}) that:
\begin{eqnarray}
\int_{-\infty}^{\infty}D\xi \cosh\sqrt{\Delta_t}=2,
\\
\beta_t g_t=6
\label{tric1}
\end{eqnarray}
where $\Delta_t$ is defined similar to $\Delta$ above with: 
\begin{equation}
\bar{\chi}_{0t}=\frac{1}{3}\int_{-\infty}^{\infty}D\xi
\frac{\sinh{\sqrt{\Delta_t}}}{\sqrt{\Delta_t}}
\label{tric2},
\end{equation} 
where the sub-index $t$ stands for the tricritical values of the temperature $T$, $g$ and $\Gamma$. 
We have solved numerically Eqs. (\ref{tric1}) and (\ref{tric2}) for several values of $\Gamma_t$. 
The result are shown in the phase diagrams (see Figs. (\ref{fig1}), (\ref{fig2}) and (3)).

\section{Results}
The numerical solutions of Eqs. (\ref{eta})-(\ref{chi}) allow one to construct two sorts of phase diagrams when 
$\Gamma$ and $g$ are independent parameters. 
The first one is $T/J$ ($T$ is the temperature) {\it versus} $g/J$ ($g$ is the strength of the pairing interaction) 
where the transverse field $\Gamma/J$ is kept constant. The second one  is $T/J$ {\it versus} $\Gamma/J$ with 
$g$ being constant. 

In the Fig. 1, one can see the results obtained in a phase diagram 
$T/J$ {\it versus} $g/J$  for $\Gamma/J=0$ and $\Gamma/J=1$ (for numerical purposes $J=1$). 
In the first case, we have obtained the same phase diagram already found in Ref. \cite{magal99} with 
three distinct regions. The normal-paramagnetic (NORMAL) region at high $T$ and small $g$ (where $q=0$ and $\eta=0$). 
For $g>g_{1}(T)$ (see section 1) one enters in the PAIR phase (where $q=0$ and $\eta\neq 0$). Finally, for low $T$ and small $g$, 
one has the phase transition 
to the spin glass phase (SG) at $T=T_{f}$ ($T_{f}$ is the freezing temperature). 
When $\Gamma$ is turned on, the freezing temperature 
decreases and the line transition $g=g_{1}(T)$ is displaced showing a dependence with $\Gamma$. Therefore, it 
is necessary to increase simultaneously the parameter $g$ to find again solutions of the order parameters 
(see Eqs. (\ref{eta})-(\ref{chi})), which corresponds to the PAIR phase. 
The position of 
the tricritical point ($T_{tc}$, $g_{tc}$)  also moves when $\Gamma$ is increased,  and 
the multiple solutions region corresponding to the first order line transition is enhanced. We also have obtained 
the behavior of the Almeida-Thouless (AT) 
eigenvalue $\lambda_{AT}$ showing that for both values of $\Gamma$, the replica symmetric SG solution is unstable. 

In Fig. 2, the phase diagram is plotted $T/J$ against $\Gamma/J$ for several values of $g$. For $g=0$ (see Fig 2.a), 
the corresponding phase diagram reproduces basically  
the results found in Ref. \cite{Alba2}. These results show that the freezing temperature $T_{f}$ decreases (when $\Gamma$ increases) 
towards to a QCP with $\Gamma_{c}=2\sqrt{2}$. The entire SG region in the phase diagram is  unstable (see the A-T line in Fig. (2.a)). 
If $g$ is turned on, which energetically favors the double occupation, the PAIR 
phase starts to appear. The existence of solution where $q=0$ and $\eta\neq 0$ depends on the ratio 
$\Gamma/g$.
For instance, 
in Fig. 2b, one can find solutions for the order parameters 
which corresponds to the PAIR phase only for small values of 
$\Gamma$. In Fig. 2c and Fig. 2d, the strength $g$ is increased and solutions 
with $\eta\neq 0$ starts to appear in a larger region of the diagram. 
Therefore, the sequence Fig. 2a-Fig. 2d shows clearly that it is requested greater 
values of $g$  to the PAIR phase occupy a larger region than the SG phase. 
The QCP given by $\Gamma=\Gamma_{c}$ is the same as $g=0$ and the SG phase remains unstable according to 
the calculated A-T line.  

Howewer, one interesting effect in the interplay between SG and the PAIR phase can be seen if one considers that 
$\Gamma$ and $g$ 
are no longer independent parameters.  
Assuming
the following relationship:
\begin{eqnarray}
\Gamma= \alpha g + \Gamma_{0}  
\label{loco}
\end{eqnarray} 
As a justification of Eq. (\ref{loco}), one should
recall the derivation of the effective model  (see Eq (\ref{ham})) given in the Appendix of the Ref. \cite{magal99}. 
The $s$-$d$ exchange part, after the integration of the conducting electrons, 
originates the pairing interaction as well as the RKKY coupling between the localized spins. 

The results have been shown in Fig. (3) in a diagram $T/J$ {\it versus} $g/J$. 
The position of the QCP ($g_{c}$) and the tricritical point ($T_{tc}$, $g_{tc}$) is very sensitive to the 
choice of the factors $\alpha$ and $\Gamma_{0}$. These factors have been adjusted to obtain  a second order transition
between the NORMAL and PAIR phases with the tricritical point located in the same scale of the Figs. (1-2).  
Therefore, for $\alpha=0.0903$ and $\Gamma_{0}=1.9254$, one can see that as long as $g$ increases 
($\Gamma$ also start to increase), 
the results show the freezing 
temperature $T_{f}$ being depressed to zero at $g_{c}$. 
For $g>g_{c}$, the solutions for the order parameters indicate a NORMAL phase until one gets
a line transition $g_{1}(T)$  
between the NORMAL and the PAIR phases  and $g_{tc}>g_{c}$. 
This phase diagram build with only one independent parameter 
($g$) is more adequate to address the experiments.

\section{Conclusions} 

In this work, we have investigated the competition between  pairing formation in real space and spin glass order 
when tunneling is tuned by the transverse field $\Gamma$. We used the same 
framework of Refs. \cite{magal99,magal00,Alba2}, 
therefore, the partition function is obtained using the functional integral formalism and the spins operators are 
represented by Grassmann variables. 
One important point is the use of the static approximation and the symmetry replica 
``ansatz" in our approach. It is known that treatment which neglects the dynamical behavior of correlation functions 
is not correct at low temperatures. Nevertheless, our interest is mainly to capture the effects  
that appears on the phase boundaries when quantum tunneling is present due to the transverse field $\Gamma$. This
procedure to find the phase boundaries 
is justified by the critical behaviour
of quantum rotor model \cite{Sachato}. 

The main results can be seen in Figs (1), (2) and (3). The first two figures
show that the pairing 
formation is not favored when the quantum tunneling is increased. At the same time, the temperature where is 
found the non-trivial spin glass ergodicity breaking decreases toward zero.  
For instance, in the Fig. 1, for the case $\Gamma=0$  where it is 
found a SG phase, for small $g$ (pairing interaction strength), 
below the freezing temperature $T_{f}=0.95J$. The presence of transverse field ($\Gamma=J$) would favor the spin flipping 
destroying the double occupation of the sites. Therefore, it is necessary to increase $g$ to find 
solutions for the order parameters 
where there is pairing long range order which corresponds to the PAIR phase. 
In that sense, the transverse field inhibit the pairing formation  which makes the sites insensitives to a magnetic interaction. 
This results can be better seen in Fig. (2) which show clearly that in order to found a PAIR phase when $\Gamma$ is 
increased it is necessary greater values of the parameter $g$. 
The position of the tricritical point found in the PAIR line transition is 
quite sensitive to the presence of the transverse field. It moves up with $\Gamma$ 
enlarging the first order transition region in the phase diagram. 

In the Fig. 3, it is assumed a 
linear relationship between $\Gamma$ and 
$g$ (the strength of the pairing interaction) given in Eq (\ref{loco}). 
Therefore, the strength of spin flipping is now related with the strength of the pairing interaction.   
he first effect when $g$ is increased is to lead the boundary line PARA-SG to a QCP at $g_{c}$. The pair formation is still
inhibited even if $g$ is kept increasing for values greater than $g_{c}$. 
The PAIR phase only appears at  the line transition $g=g_{1}(T)$. This resulting phase diagram displays 
phase boundaries similar to the experimental one for  $U_{1-x}$$La_{x}$$Pd_{2}Al_{3}$ when 
$x>0.5$ \cite{1}.   
 
\begin{figure}[t]
\includegraphics[angle=270,width=12cm]{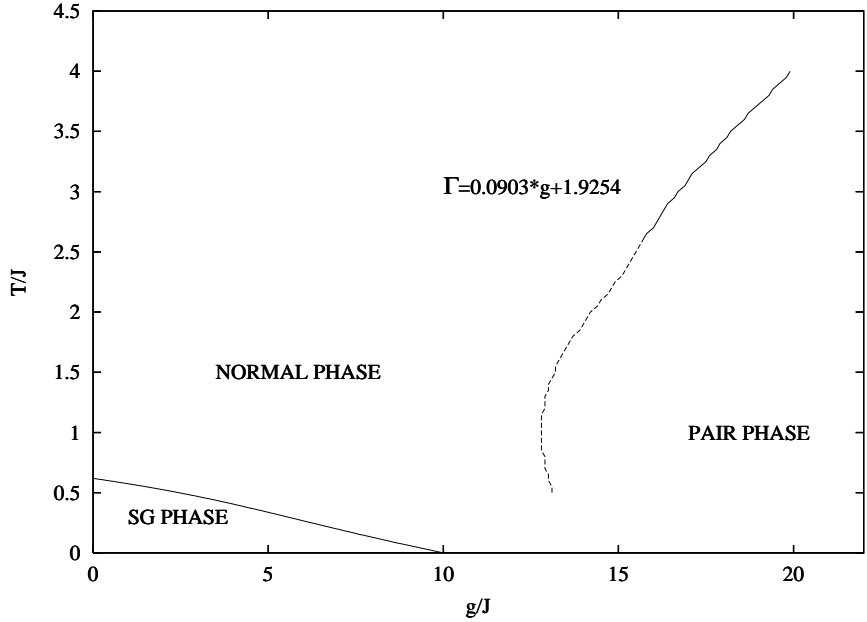}
\caption{Phase diagram build with a relationship between $\Gamma$ and $g$. The dotted line indicates the first 
order transition while solid line the second order transition.}  
\label{fig3}
\end{figure}

In conclusion, we have studied a fermionic representation of Ising spin glass (SG) in the presence of 
transverse magnetic field 
$\Gamma$ together with local pairing 
interaction.    
We expect that results obtained in this model can 
contribute for the study of the interplay between spin glass and superconductivity in 
strongly correlated systems. Particularly, to describe the phase boundaries which is the 
main interest of this work. 
It should remarked that we have used  the replica symmetry ansatz in the present work. 
There are  results \cite{Oppermann1} indicating 
the Parisi replica permutation symmetry breaking 
affect the boundary between superconductivity and the SG phase.   
This is an indication that 
would be necessary to go beyond the replica symmetry solution in the present work. 
That will be subject for future work.         

\vspace{0.5cm}
{\bf Acknowledgement}

The numerical calculations were partially performed at  LSC (Curso de Ci\^encia da Computa\c{c}\~ao, UFSM). 
SGM is grateful to Prof. Alba Theumann for useful comments. 
This work was partially supported by the Brazilian agencies FAPERGS (Funda\c{c}\~ao de Amparo \`a 
Pesquisa do Rio Grande do Sul) 
and CAPES (Coodena\c{c}\~ao de Aperfei\c{c}oamento de Pessoal de Nivel Superior).

\end{document}